\documentclass[aps,reprint,superscriptaddress]{revtex4-2}
\usepackage{amsmath}
\usepackage{amssymb}
\usepackage{graphicx}
\usepackage{orcidlink}
\usepackage{hyperref}
\usepackage{bm}
\usepackage{color}
\hypersetup{colorlinks=true}

\begin{document}


\title{From linear to nonlinear Breit-Wheeler pair production in laser-solid interactions}

\author{Huai-Hang Song\,\orcidlink{0000-0002-2587-4658}}\email{huaihangsong@sjtu.edu.cn}
\affiliation{Key Laboratory for Laser Plasmas (MOE) and School of Physics and Astronomy, Shanghai Jiao Tong University, Shanghai 200240, China}
\affiliation{Collaborative Innovation Center of IFSA, Shanghai Jiao Tong University, Shanghai 200240, China}

\author{Wei-Min Wang\,\orcidlink{0000-0002-9852-1589}}
\affiliation{Department of Physics and Beijing Key Laboratory of Opto-electronic Functional Materials and Micro–nano Devices, Renmin University of China, Beijing 100872, China}
\affiliation{Collaborative Innovation Center of IFSA, Shanghai Jiao Tong University, Shanghai 200240, China}

\author{Min Chen\,\orcidlink{0000-0002-4290-9330}}
\affiliation{Key Laboratory for Laser Plasmas (MOE) and School of Physics and Astronomy, Shanghai Jiao Tong University, Shanghai 200240, China}
\affiliation{Collaborative Innovation Center of IFSA, Shanghai Jiao Tong University, Shanghai 200240, China}

\author{Zheng-Ming Sheng\,\orcidlink{0000-0002-8823-9993}}\email{zmsheng@sjtu.edu.cn}
\affiliation{Key Laboratory for Laser Plasmas (MOE) and School of Physics and Astronomy, Shanghai Jiao Tong University, Shanghai 200240, China}
\affiliation{Collaborative Innovation Center of IFSA, Shanghai Jiao Tong University, Shanghai 200240, China}
\affiliation{Tsung-Dao Lee Institute, Shanghai Jiao Tong University, Shanghai 201210, China}

\date{\today}

\begin{abstract}

During the ultraintense laser interaction with solids (overdense plasmas), the competition between two possible quantum electrodynamics (QED) mechanisms responsible for $e^\pm$ pair production, i.e., linear and nonlinear Breit-Wheeler (BW) processes, remains to be studied.  Here, we have implemented the linear BW process via a Monte Carlo algorithm into the QED particle-in-cell (PIC) code {\scshape yunic}, enabling us to self-consistently investigate both pair production mechanisms in the plasma environment. By a series of 2D QED-PIC simulations, the transition from the linear to the nonlinear BW process is observed with the increase of laser intensities in the typical configuration of a linearly polarized laser interaction with solid targets. A critical normalized laser amplitude about $a_0\sim 400$--500 is found under a large range of preplasma scale lengths, below which the linear BW process dominates over the nonlinear BW process. This work provides a practicable technique to model linear QED processes via integrated QED-PIC simulations. Moreover, it calls for more attention to be paid to linear BW pair production in near future 10-PW-class laser-solid interactions. 

\end{abstract}


\maketitle


\section{Introduction}

Since the invention of chirped pulse amplification \cite{strickland1985oc} and other relevant technologies \cite{li2023lpr}, the peak intensity of fs lasers has been increased by 7 to 8 orders of magnitude in the last three and a half decades \cite{danson2019hplse}. The strongest laser intensity available now is claimed to reach $10^{23}$ W/cm$^2$ by near diffraction-limited focusing the CoReLS 4 PW laser to a spot size of 1.1 $\mu$m \cite{yoon2021optica}. Several 10 PW laser facilities are completed or near completion in different places in the world, such as SULF \cite{li2022hplse}, ELI-NP \cite{tanaka2020mre}, Apollon \cite{burdonov2021mre}. Besides, a 100 PW laser facility SEL has been under construction \citep{shao2020ol}. The peak intensity of these $10$--$100$ PW lasers when well-focused is expected to be $10^{23}$--$10^{24}$ W/cm$^2$, i.e., with the normalized amplitude $a_0=eE_0/m_ec\omega_0\sim 300$--800, where $E_0$ and $\omega_0$ are the electric field amplitude and frequency of the laser, $e$ is the elementary charge, $m_e$ is the rest mass of electron, and $c$ is the speed of light in vacuum.

The increase of laser intensities above $10^{22}$ W/cm$^2$ will bring the laser-plasma interaction from the relativistic regime \cite{gibbon2005book} to the strong-field quantum electrodynamics (QED) regime \cite{piazza2012rmp, gonoskov2022rmp, fedotov2023pr}, in which two dominated phenomena are $\gamma$ photon emission and $e^\pm$ pair production. The most efficient mechanism responsible for photon emission is often nonlinear Compton scattering that energetic electrons are scattered by the strong field and emit $\gamma$ photons ($e+n\omega_0=e'+\omega$) \cite{bula1996prl}. The subsequent decay of $\gamma$ photons into $e^\pm$ pairs has relatively richer mechanisms \cite{hubbell2006rpc, ruffini2010pr}, such as linear Breit-Wheeler (BW) process \cite{breit1934pr}, nonlinear BW process \cite{reiss1962jmp}, Bethe-Heitler process \cite{bethe1034prsa}, triplet pair production \cite{joseph1958rmp} and so on. In this paper, we restrict the subject to linear and nonlinear BW processes.

As the inverse process of $e^\pm$ annihilation into photons \cite{dirac1930mpcps}, the linear BW process refers to the collision of two photons to produce a pair of $e^\pm$ when the photon energies reach a certain threshold ($\omega_1+\omega_2=e^-+e^+$). As one of the most fundamental QED processes, although it has been confirmed by some experiments \cite{baur2007pr,adam2021prl}, its direct observation by pure photons is not realized yet. There are many proposals to detect linear BW pair production, e.g., by the aid of brilliant plasma-based photon sources \cite{ribeyre2016pre,yu2019prl,wang2020prl} or laser-heated hohlraums \cite{pike2014np}.

The nonlinear BW process is related to the decay of a $\gamma$ photon into a pair of $e^\pm$ in the strong field ($\omega+n\omega_0=e^-+e^+$). It has been experimentally demonstrated in 1990s by combining the traditional accelerator and a TW laser at SLAC facility \cite{burke1997prl}. Over the past decade, the nonlinear BW process has been extensively studied in both laser-electron collisions \cite{piazza2016prl,blackburn2017pra} and laser-plasma interactions \cite{nerush2011prl,ridgers2012prl}. The latter is facilitated by the fact that currently most of PIC codes have the ability to simulate nonlinear Compton scattering and nonlinear BW process based on the locally constant field approximation (LCFA) \cite{ritus1985jslr, baier1998qed}, such as widely used open-source {\scshape epoch} \cite{arber2015ppcf} and {\scshape smilei} \cite{derouillat2018cpc} codes.

In comparison, the linear BW process is paid much less attention in the laser-plasma community. There are two possible reasons for this dilemma. One reason comes from the misconception that the linear BW process was considered unimportant because of its low cross section, unless two brilliant photon beams are first produced by two laser pulses with plasmas and then collided \cite{yu2019prl,wang2020prl,he2021cp}. However, some recent works have indicated that the linear BW process will become the dominated mechanism even in the single laser interaction with microchannel targets \cite{he2021njp} or near-critical targets \cite{sugimoto2023prl}. Another reason is that the QED-PIC code is considered to be unable to handle linear QED processes for a long time and one has to rely on some post-processing methods. Until recently, a linear BW algorithm is implemented into a PIC code \cite{sugimoto2023prl}, where the photons are treated as rays rather than the usual microparticles.

In this paper, we present a scheme to implement a photon collision algorithm for the linear BW process into the QED-PIC code {\scshape yunic} \cite{song2021arxiv}. The adopted algorithm was originally designed for nuclear fusion \cite{higginson2019jcp} but can be easily extended to other collision-related reactions. It is particularly suitable for inhomogeneous plasmas because of different weights of microparticles usually in the modern PIC structure. Using our newly developed QED-PIC code, we can self-consistently investigate both linear and nonlinear BW processes in a more typical laser-plasma system, i.e., laser-solid interactions. The transition from linear to nonlinear BW pair production is observed with the increase of laser intensities. Dense photon emission coupled with a large divergence angle allows the linear BW process to dominate over the nonlinear BW process at low laser intensities of $a_0<400$ over different preplasma scale lengths $L_0=0.5$--$2.0~\mu$m.

This article is structured as follows. Section \ref{algorithm} introduces the adopted algorithm for the linear BW process and provides code benchmarks. In Sec.~\ref{results}, we present our simulation setup and results in typical laser-solid interactions. Linear and nonlinear BW processes are analyzed, and their dominance is compared at different laser intensities and preplasma scale lengths. A brief conclusion is drawn in Sec. \ref{conclusion}.

\section{Implementing linear BW process into QED-PIC code}
\label{algorithm}

\subsection{Algorithm introduction}

The cross section of the linear BW process after averaging the polarization effect is widely known as \cite{breit1934pr, jauch1976book}
\begin{equation}\label{eq1}
\sigma_{\gamma\gamma}=\frac{\pi r_e^2}{2}(1-\nu^2)\left[2\nu^3-4\nu+(3-\nu^4)\ln(\frac{1+\nu}{1-\nu})\right],
\end{equation}
where $\nu=\sqrt{1-1/\mu}$, $\mu=\varepsilon_1\varepsilon_2(1-\cos\theta)/2m_e^2c^4$, and other parameters are photon energy $\varepsilon_{1,2}$, collision angle $\theta$, and classical electron radius $r_e=e^2/m_ec^2\approx2.82\times10^{-13}$ cm.

Our Monte Carlo algorithm basically follows the idea of Ref.~\cite{higginson2019jcp}, which can be easily extended from original nuclear fusion to linear BW pair production because they both essentially use the pairwise collision of microparticles. In the following, we introduce it step by step.

(I) \emph{Randomly pairing microphotons in the same cell}. Like Coulomb collision \cite{takizuka1977jcp} and nuclear fusion \cite{higginson2019jcp} algorithms, the collision associated with the linear BW process in the PIC code only occurs for microphotons located in the same cell. This basically does not cause errors because the microparticles in neighboring cells are statistically similar. One should record the address of one-cell microphotons and also the corresponding microphoton number $\mathbb{N}_\gamma$ in each cell. Next, the order of microphotons in the same cell should be shuffled before pairing. If $\mathbb{N}_\gamma$ is an even number, all microphotons can be assigned in pairs, with the total collisions $\mathbb{N}_\gamma/2$; if $\mathbb{N}_\gamma$ is an odd number, the last microphoton will not participate in the collision, with the total collisions $(\mathbb{N}_\gamma-1)/2$.

(II) \emph{Calculating the reaction probability}. For the collision between two microphotons of weights $w_1$ and $w_2$, the reaction yield can be expressed as
\begin{equation}\label{eq2}
Y_{\gamma\gamma}=\mathbb{N}_{\rm ratio} w_1w_2\sigma_{\gamma\gamma}(1-\cos\theta)c\Delta t/\Delta V,
\end{equation}
where $\Delta t$ is the numerical time-step and $\Delta V$ is the cell volume in the PIC code. The factor $\mathbb{N}_{\rm ratio}$ is to compensate for not making all possible pairings. In principal, arbitrary two microphotons in one cell should be paired with the total collisions $\mathbb{N}_\gamma(\mathbb{N}_\gamma-1)/2$, but actually we only pair them once as described in step (I). Therefore, the compensation factor is $\mathbb{N}_{\rm ratio}=\mathbb{N}_\gamma-1$ for the even $\mathbb{N}_\gamma$, while $\mathbb{N}_{\rm ratio}=\mathbb{N}_\gamma$ for the odd $\mathbb{N}_\gamma$.

The centre idea of the adopted algorithm \cite{higginson2019jcp} is not to directly use Eq.~\eqref{eq2}, while the following modified equation for the reaction probability
\begin{equation}\label{eq3}
P_{\gamma\gamma}=\mathbb{F}_{\rm mult}\mathbb{N}_{\rm ratio} {\rm Max}(w_1,w_2)\sigma_{\gamma\gamma}(1-\cos\theta)c\Delta t/\Delta V.
\end{equation}
The original Eq.~\eqref{eq2} is enlarged by a multiplication factor $\mathbb{F}_{\rm mult}\gg1$, but at the same time one should ensure that $P_{\gamma\gamma}<1$.

(III) \emph{Creating $e^\pm$ pairs in the center-of-mass frame}. If the probability $P_{\gamma\gamma}$ is greater than a random number between 0 and 1, then a microelectron and a micropositron with the same weight
\begin{equation}\label{eq4}
w_+=w_-={\rm Min}(w_1,w_2)/\mathbb{F}_{\rm mult}
\end{equation}
are created. Simultaneously, the weights of two involved microphotons are both accordingly reduced by $w_\pm$. Obviously, the weights of the created microelectron and micropositron are $\mathbb{F}_{\rm mult}$ times less than the minimum weight of their two parent microphotons. One can easily check that the $e^\pm$ yields calculated from Eq.~\eqref{eq2} and Eqs.~\eqref{eq3}-\eqref{eq4} are the same by checking $Y_{\gamma\gamma}=P_{\gamma\gamma}w_\pm$.

Introducing the multiplication factor $\mathbb{F}_{\rm mult}$ in Eq.~\eqref{eq3} holds two advantages. On the one hand, the parts of two microphotons actually involved in the collision have the same weight, so the created microelectron and micropositron also have the same weight [see Eq.~\eqref{eq4}]. This step can make energy and momentum conservation more easily before and after the collision \cite{higginson2019jcp}. Another advantage is that more microelectrons and micropositrons can be created, indicating a better statistical result with the limited initial microparticles in the PIC simulations. Of course, the actual $e^{\pm}$ yield is not influenced because the weights of created microelectrons of micropositrons are decreased accordingly. 

The energy and momentum of newly created $e^\pm$ pairs are usually first calculated in the center-of-mass (COM) frame. The velocity of the COM frame is $\bm v_{\rm c}=(\bm p_1+\bm p_2)c^2/(\varepsilon_1+\varepsilon_2)$, where $\bm p_{1,2}$ is the photon momentum. The energy of the created $e^\pm$ in the COM frame is $\varepsilon'_\pm=\frac{1}{2}\sqrt{(\varepsilon_1+\varepsilon_2)^2-(\bm p_1c+\bm p_2c)^2}$, and the corresponding $e^\pm$ momentum is $p'_\pm=\frac{1}{c}\sqrt{{\varepsilon'}_\pm^2-m_e^2c^4}$. The direction of the $e^\pm$ momentum is isotropically selected as long as $\bm p'_-=-\bm p'_+$ in the COM frame.

(4) \emph{Transforming back to the laboratory frame}.
The $e^\pm$ momentum obtained in the COM frame should be finally transformed into the laboratory frame via Lorentz transformation $\bm p_\pm=\bm p'_\pm+(\gamma_c-1)(\bm p'_\pm\cdot \bm e_c)\bm e_c-\gamma_c{\varepsilon'}_\pm\bm v_c/c$, where $\bm e_c=\bm v_c/v_c$, $\gamma_c=1/\sqrt{1-v_c^2/c^2}$.

\subsection{Code benckmarks in the high-temperature blackbody radiation}

\begin{figure}[t]
\centering
\includegraphics[width=\columnwidth]{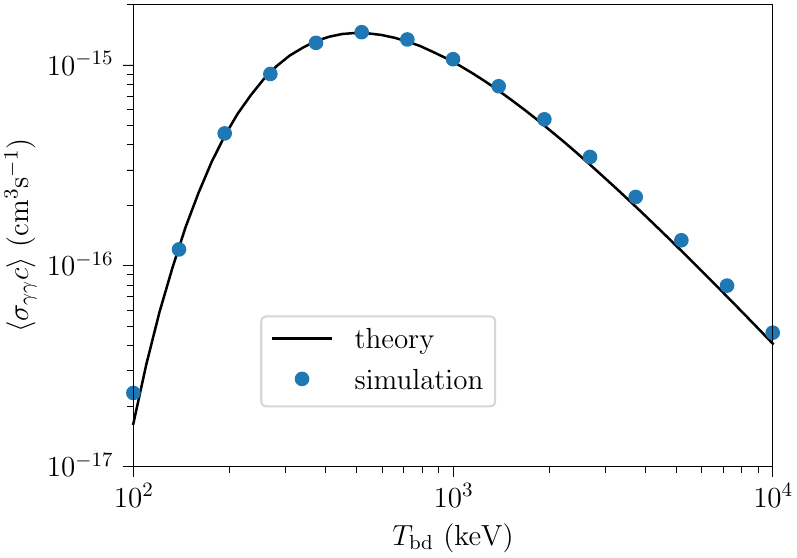}
\caption{\label{fig1} Benchmarks of our simulation results with the theory of the linear BW pair production rate $\langle\sigma_{\gamma\gamma}c\rangle$ versus the temperature $T_{\rm bd}$ in the high-temperature blackbody radiation.}
\end{figure}

Here, we benchmark above Monte Carlo algorithm by simulating linear BW pair production in the high-temperature blackbody radiation \cite{weaver1976pra,beesley2022rp}. The photon density of the blackbody radiation obeys the Planck distribution $n_{\rm bd}=(2\omega^2/\pi c^3)/[\exp(\hbar\omega/k_BT_{\rm bd})-1]$, where $\hbar$ is the reduced Planck constant and $k_B$ is the Boltzmann constant. If the blackbody temperature $T_{\rm bd}$ is close to the electron rest energy $m_ec^2$ of about 511 keV, a significant portion of photons will be converted into $e^\pm$ pairs by stochastic collisions via the linear BW process. Such a high blackbody temperature could be reached in early universe shortly after Big Bang (lepton epoch). Its pair production rate has an analytic formula $\langle\sigma_{\gamma\gamma}c\rangle=2R_{\gamma\gamma}/n_{\rm bd}^2$, where the specific expression of $R_{\gamma\gamma}$ can be found in Eq.~(30) of \cite{weaver1976pra}. We simulate such an extreme process using the 1D-version {\scshape yunic} code \cite{song2021arxiv} with the blackbody temperature ranging from $10^2$ keV to $10^4$ keV. The simulation box length is 10$^{-6}~\mu$m and resolved by 192 cells. In each cell, 500 microphotons are filled. The periodic boundary is used and the multiplication factor $\mathbb{F}_{\rm mult}$ is set to 100.

The pair production rate $\langle\sigma_{\gamma\gamma}c\rangle$ as a function of the blackbody temperature $T_{\rm bd}$ is shown in Fig.~\ref{fig1}. Our simulation results are well consistent with the theoretical prediction. We have also conducted another set of simulations with the same parameters except that each cell is filled with a random microphoton number between 100 and 900. No obvious difference is observed, indicating our algorithm can correctly handle the microphoton collision of different weights.

\section{Simulation Results and analysis}
\label{results}

\subsection{Simulation setups}

After code benchmarks, a series of QED-PIC simulations by the 2D-version {\scshape yunic} code \cite{song2021arxiv} are carried out to investigate both linear and nonlinear BW processes during the laser-solid interaction. The algorithm of linear BW pair production in our code has been introduced in Sec.~\ref{algorithm}, and the nonlinear BW algorithm based on LCFA was implemented and benchmarked before \cite{song2021arxiv}.

In the typical simulation case, a fully ionized solid target has a thickness $5~\mu$m and a bulk electron density $200n_c$, where $n_c=m_e\omega_0^2/2\pi e^2$ is the critical plasma density. In the front of the target, there are some low-density preplasmas of a scale length $L_0=1~\mu$m, considering that in real experiments low-intensity prepulses are hard to avoid. A laser pulse linearly polarized along the $y$ direction is incident from the left boundary ($x=0$) to vertically impinge on the solid target. The laser pulse has a centre wavelength $\lambda_0=1~\mu$m, a normalized amplitude $a_0=400$, a FWHM duration $16.7$ fs, and a waist radius $3~\mu$m. The simulation domain is $L_x\times L_y=15~\mu\rm m\times12~\mu\rm m$, evenly resolved by $480\times384$ cells. In each cell, 9 microelectrons and 9 microproton are filled. Absorbing boundaries are taken for both fields and particles in each direction. The 4th-order interpolation for the microparticle shaping is adopted to reduce numerical errors.

\subsection{Typical simulation results}
\label{typical}

\begin{figure*}[]
\centering
\includegraphics[width=\textwidth]{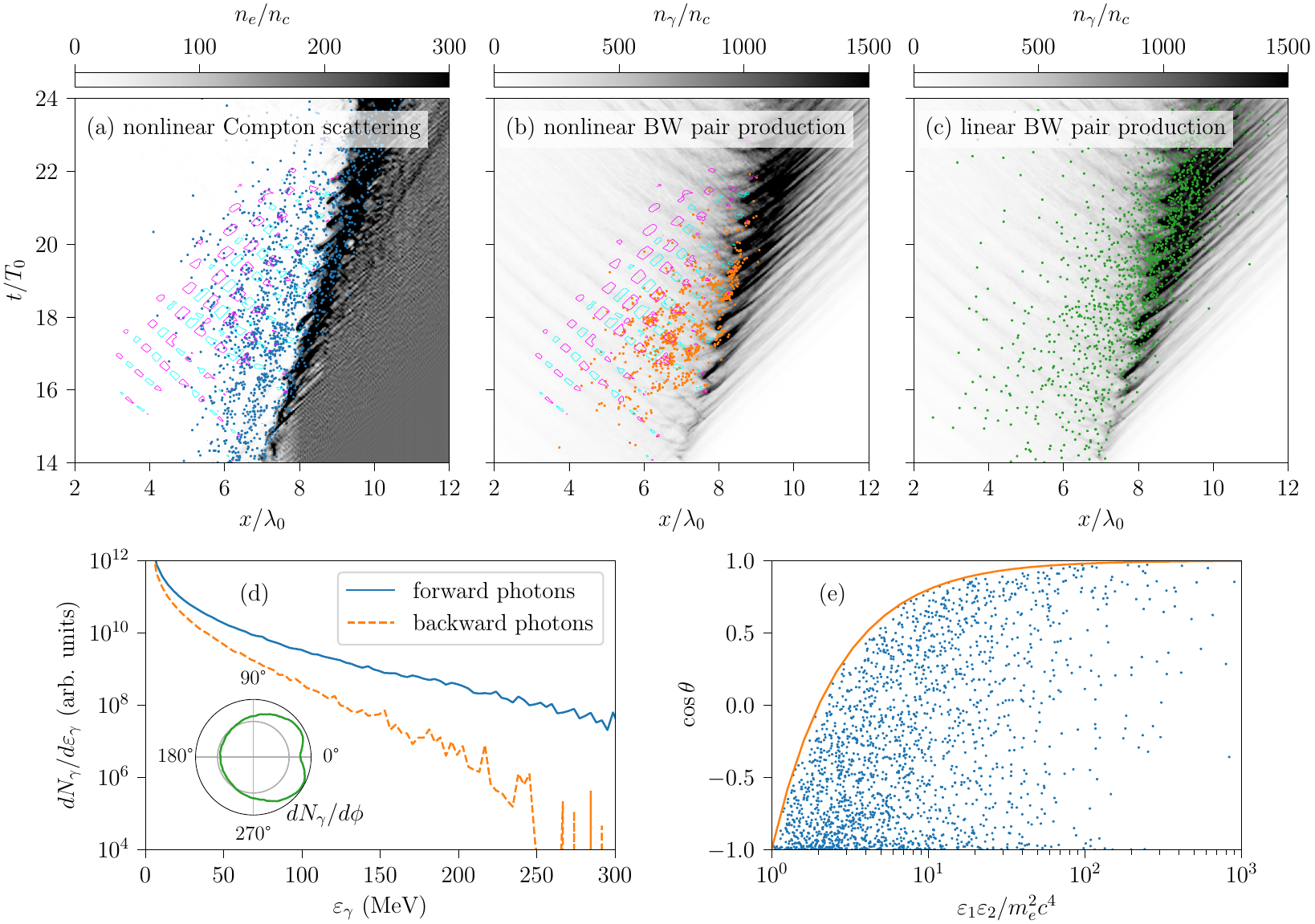}
\caption{\label{fig2} Spatiotemporal distributions of (a) electron density $n_e$ and some nonlinear Compton scattering events (blue dots), (b) photon density $n_\gamma$ and some nonlinear BW pair production events (orange dots), (c) photon density $n_\gamma$ and some linear BW pair production events (green dots). Contour lines of $e(|B_z|-|E_y|)/m_ec\omega_0=300$  (magenta lines) and $-300$ (cyan lines) in (a) and (b) are also plotted to distinguish MFDRs and EFDRs, respectively. (d) The spectra of forward photons and backward photons, and the insert shows the angular number distribution of emitted photons. (e) The quantities $\varepsilon_1\varepsilon_2/m_e^2c^4$ versus $\cos\theta$ of some decayed photons during linear BW pair production.}
\end{figure*}


In Figs.~\ref{fig2}(a)-\ref{fig2}(c), spatiotemporal distributions of electron density and photon density, together with some randomly selected QED events that occurred within 0.5 $\mu$m of the laser axis are shown, corresponding to nonlinear Compton scattering, nonlinear BW process and linear BW process, respectively. Let us elaborate on them one by one below.

\emph{Nonlinear Compton scattering}. For both linear and nonlinear BW processes, photon emission via nonlinear Compton scattering is the first step. In Fig.~\ref{fig2}(a), some photon emission events are recorded in the background of electron density. It can be seen that photons are mainly emitted in the front of the target surface, where the electromagnetic standing wave formed by the incident and reflected laser pulses can greatly favor nonlinear strong-field QED processes \cite{bashmakov2014pop,esirkepov2015pla}. In the case of the linearly polarized laser pulse, electrons can be effectively accelerated in the electric-field-dominated region (EFDR, $|B_z|<|E_y|$) and a large QED parameter $\chi_e=(e\hbar/m_e^3c^4)|F_{\mu\nu}p^{\nu}|$ that measures the strength of photon emission is achieved in the magnetic-field-dominated region (MFDR, $|B_z|>|E_y|$), where $F_{\mu\nu}$ is the field tensor and $p^{\nu}$ is the electron four-momentum. These two spatiotemporally separated regions mutually contribute to nonlinear QED processes. This is verified by our simulation that the ratio of photon emission events happened in the MFDR and EFDR is about 2.1. 

\emph{Nonlinear BW pair production}. The nonlinear BW process is very similar to nonlinear Compton scattering, with symmetric Feynman diagrams and analogous probability expressions \cite{bulanov2013pra}, both of which rely on the transverse components of the strong field. The emitted photons via nonlinear Compton scattering can further decay into $e^\pm$ pairs via the nonlinear BW process in the electromagnetic standing wave, as shown in Fig.~\ref{fig2}(b). Similarly, nonlinear BW pairs are also mainly created in the MFDR, with a ratio of about 4.1 to the EFDR in our simulation. However, different from nonlinear Compton scattering, the nonlinear BW process is exponentially suppressed in the weak QED regime. The pair production probability can be approximately written as $0.23\Delta t(\alpha m_e^2c^4/\hbar\varepsilon_\gamma)\chi_\gamma\exp(-8/3\chi_\gamma)$ in the low $\chi_\gamma$ limit, where $\chi_\gamma=(e\hbar^2/m_e^3c^4)|F_{\mu\nu}k^{\nu}|$ measures the strength of the nonlinear BW process,
with $\hbar k^{\nu}$ being the photon four-momentum and $\alpha=e^2/\hbar c\approx 1/137$ being the fine structure constant. This exponential probability has an error of less than 30\% compared to the exact value for $\chi_\gamma<0.5$. Let us estimate the decay probability of photons into $e^\pm$ pairs via the nonlinear BW process by simply considering a photon counterpropagating in the laser field of the normalized laser amplitude $a_0=400$ and an interaction duration of 1 fs. When the photon energy is set to 100 MeV, the photon decay probability is about $10^{-3}$ by taking the maximum $\chi_\gamma\approx0.38$; however, if the photon energy is decreased to 30 MeV, the corresponding probability is dramatically decreased to only about $10^{-10}$ by taking the maximum $\chi_\gamma\approx0.11$. In the laser-plasma interaction, most of photons are low-energy, resulting in an overall decay probability of about $3\times10^{-6}$ in our simulation. Since the laser intensity controls photon energies and consequently $\chi_\gamma$, it is not difficult to infer that the nonlinear BW process is highly dependent on the laser intensity, especially at the weak laser intensity.

\emph{Linear BW pair production}. By comparing Figs.~\ref{fig2}(b) and \ref{fig2}(c), one obvious feature is that the linear BW process mainly happens near the target surface, rather than in the target front. This is because the linear BW process is directly attributed to the collision of photons, not laser fields. Therefore, the photon density and the emission angle are two crucial factors for the occurrence of photon collisions. Fortunately, irradiating the solid target with even a single laser pulse, there are dense photons emitted with large divergence angles, as shown in Figs.~\ref{fig2}(c) and \ref{fig2}(d). About 2/3 photons are emitted forward but still 1/3 photons are emitted backward, providing a sufficient condition for the photon collision. Besides, the forward photons have a higher temperature than backward photons, seen from Fig.~\ref{fig2}(d). We can also estimate the probability of linear BW pair production by simplifying the complex laser-plasma configuration into a head-on collision of two photon beams of the same other parameters. The photon density and beam length are taken to $n_\gamma=1000n_c$ and $\ell=1~\mu$m, similar to those observed from Fig.~\ref{fig2}(c). The decay probability of photons into $e^\pm$ pairs via the linear BW process is about $2\times10^{-9}\times\sigma(10^{-18}~\mu{\rm m}^2)\times \ell(\mu{\rm m})\times n_\gamma/n_c\approx 4\times10^{-6}$ if a fixed cross section $\sigma_{\gamma\gamma}=2\times10^{-18}~\mu{\rm m}^2$ is taken. The estimated probability is close to our simulation of about $3\times10^{-6}$. In the actual situation, both collision angles and energies of participating photons have wide ranges, as shown in Fig.~\ref{fig2}(e). Most of linear BW events happen near the head-on collision of $\cos\theta\rightarrow-1$. Due to the very high energy of photons, a considerable proportion of photons can even decay into pairs in the near copropagating case of $\cos\theta\rightarrow1$ as long as $\varepsilon_1\varepsilon_2(1-\cos\theta)>2m_e^2c^4$.

\subsection{Transition from linear to nonlinear BW process with the increase of laser intensities}

\begin{figure}[t]
\centering
\includegraphics[width=\columnwidth]{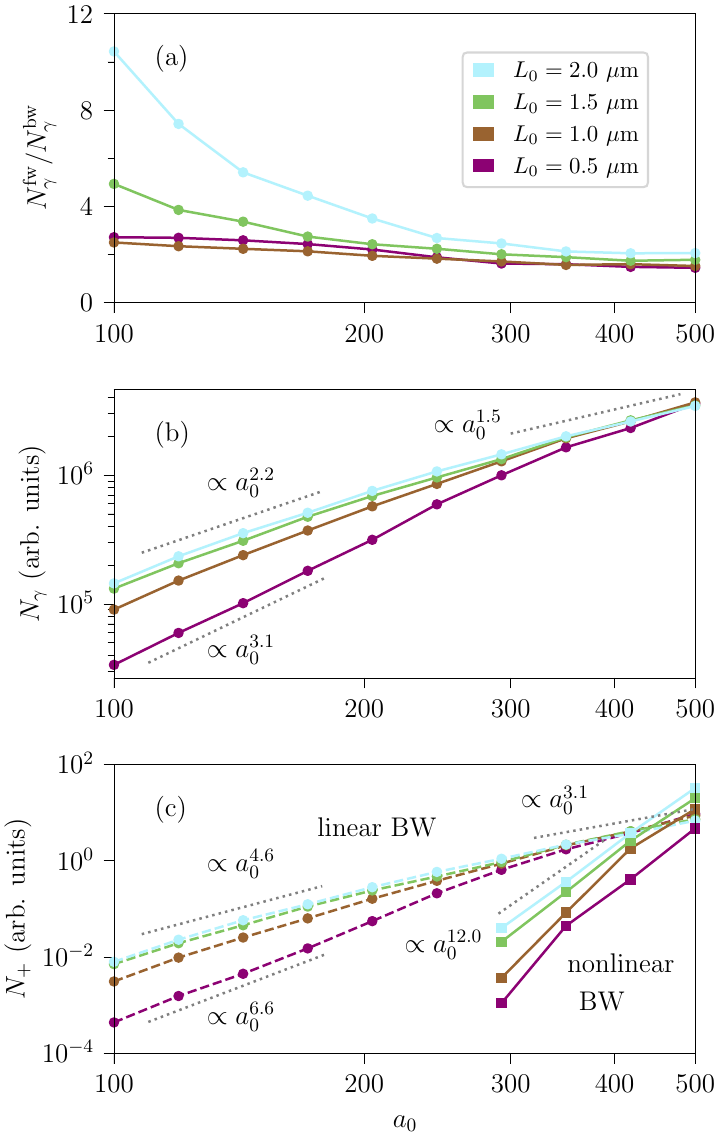}
\caption{\label{fig3} (a) The number ratio of forward photons to backward photons $N_\gamma^{\rm fw}/N_\gamma^{\rm bw}$. (b) The photon yield and (c) pair production yield $N_+$ as the function of the laser intensity $a_0$ under four preplasma scale length $L_0=0.5~\mu$m, $1.0~\mu$m, $1.5~\mu$m, and $2.0~\mu$m. In (c), the pair production yield of linear  and nonlinear BW processes corresponds to dashed-circle and solid-square curves, respectively.}
\end{figure}

Through the estimates in Sec.~\ref{typical}, the pair production rate of the linear BW process mainly relies on the photon density and emission angle, while that of the nonlinear BW process strongly relies on the QED parameter $\chi_\gamma$, especially in the weak QED regime. For the weak laser intensity, nonlinear BW pair production is strongly suppressed. In this case, one can expect that linear BW pair production will dominate. In this subsection, we will investigate the influence of the preplasma scale length ($L_0=0.5$--$2.0~\mu$m) and the laser intensity ($a_0=100$--500) on both mechanisms.

From Fig.~\ref{fig3}(a), in general, the proportion of backward emitted photons becomes larger as the laser intensity increases or the preplasma scale length decreases. For the large scale length $L_0=2.0~\mu$m and the weak laser intensity $a_0=100$, the number ratio of forward photons to backward photons $N_\gamma^{\rm fw}/N_\gamma^{\rm bw}$ can reach about 11, because the laser pulse can effectively accelerate electrons to emit photons forward and deplete most of its energy in preplasmas before reaching bulk plasmas. With the increase of laser intensities, the number ratio $N_\gamma^{\rm fw}/N_\gamma^{\rm bw}$ converges at about 2. For the small scale length $L_0<1.0~\mu$m, the ratio is not sensitive to the laser intensity, always maintaining at around 2.

At the weak laser intensity, low-density preplasmas can greatly enhance photon emission, as shown in Fig.~\ref{fig3}(b). The total photon yield can be even improved by nearly 10 times from $L_0=0.5~\mu$m to $2.0~\mu$m at $a_0=100$. With the increase of laser intensities, the boosting effect of preplasmas fades away. The scale of the photon yield to the laser intensity is decreased from $a_0^{3.1}$ to $a_0^{2.2}$ as the preplasma scale length increases from $L_0=0.5~\mu$m to $2.0~\mu$m. When the normalized laser amplitude $a_0$ exceeds 300, the corresponding scale is converged at about $a_0^{1.5}$ for all considered preplasma scale lengths.

Figure~\ref{fig3}(c) presents the positron yield from linear and nonlinear BW processes, respectively. At the weak laser intensity $a_0<200$, only linear BW pairs are presented from our simulations because nonlinear BW pairs are exponentially suppressed. Although in this weak case backward photons are relatively rare under large preplasma scale lengths, the photon yield is greatly improved, leading to that pair production via the linear BW process is also enhanced. Similarly, the degree of enhancement by preplasmas weakens with the increase of laser intensities. The final scale of the linear BW yield at high laser intensities is about $a_0^{3.1}$ for all scanned preplasma scale lengths. For all parameter ranges, the linear BW yield is approximately proportional to the square of the photon yield, which can be seen from the corresponding scales annotated in Figs.~\ref{fig3}(b) and \ref{fig3}(c). With the increase of laser intensities, one of the most features is that nonlinear BW pair production will dominate over linear BW pair production. The nonlinear BW yield is also obviously improved by preplasmas \cite{wang2017pre}. Between $a_0=300$ and 500, the local scale of the nonlinear BW yield to the laser intensity can reach $a_0^{12.0}$, and the nonlinear BW process will replace the linear one as the dominant mechanism at $a_0>400$--500 over a large range of preplasma scale lengths $L_0=0.5~\mu$m--$2.0~\mu$m.

\section{conclusion}

In conclusion, we have implemented a Monte Carlo algorithm into our QED-PIC code to self-consistently calculate linear BW pair production. The algorithm has advantages in dealing with microphoton collisions of different weights usually encountered in inhomogeneous plasmas. Our code benchmarks in the high-temperature blackbody radiation confirm its validity. Utilizing the code, we have investigated both linear and nonlinear BW processes in the typical laser-solid interaction. In general, the linear one mainly depends on the square of the photon density, while the nonlinear one exponentially depends on the photon energy or the QED parameter $\chi_\gamma$ at weak laser intensities. Thus, with the increase of laser intensities, the dominant mechanism will transition from linear to nonlinear BW pair production. The critical normalized laser amplitude for the transition between two mechanisms is about $a_0\sim400$--500. This work indicates linear BW pair production is important in the near future 10-PW-class laser-solid experiments. Meanwhile, it provides a potential application for the generation of positrons, as well as an idea for probing the linear BW process in plasma environments with a single laser. 

\label{conclusion}

\begin{acknowledgments}

This work was supported by the National Science Foundation of China (Grants  No.~12135009, 11991074 and 12225505). The simulations were performed on the $\pi$ 2.0 supercomputer at Shanghai Jiao Tong University.

\end{acknowledgments}

\bibliography{reference}

\end{document}